\documentclass[conference]{IEEEtran}
\pdfoutput=1

\usepackage[pdftex,colorlinks=false]{hyperref}
\usepackage[utf8x]{inputenc}
\usepackage{graphicx}
\usepackage{todonotes}
\usepackage{bbding}
\usepackage{array}

\usepackage{bytefield}

\newcolumntype{V}{>{\raggedleft\arraybackslash}p{1cm}}

\title{No Need for Black Chambers: Testing TLS in the E-mail Ecosystem at Large} 

\author{
    \IEEEauthorblockN{Wilfried Mayer\IEEEauthorrefmark{1},
Aaron Zauner\IEEEauthorrefmark{1},
Martin Schmiedecker\IEEEauthorrefmark{1} and Markus Huber\IEEEauthorrefmark{2}}\\
    \IEEEauthorblockA{
      \IEEEauthorrefmark{1}SBA Research, Austria\\ \textit{\{wmayer$|$azauner$|$mschmiedecker\}@sba-research.org}\\
      \IEEEauthorrefmark{2}FH St. Pölten, Austria\\ \textit{markus.huber@fhstp.ac.at}\\
    }
}

\begin{document}
\pagestyle{plain}

\maketitle

\begin{abstract}
TLS is the most widely used cryptographic protocol on the Internet. While many recent studies focused on its use in HTTPS, none so far analyzed TLS usage in e-mail related protocols, which often carry highly sensitive information. Since end-to-end encryption mechanisms like PGP are seldomly used, today confidentiality in the e-mail ecosystem is mainly based on the encryption of the transport layer. A well-positioned attacker may be able to intercept plaintext passively and at global scale.

In this paper we are the first to present a scalable methodology to assess the state of security mechanisms in the e-mail ecosystem using commodity hardware and open-source software. We draw a comprehensive picture of the current state of every e-mail related TLS configuration for the entire IPv4 range. We collected and scanned a massive data-set of 20 million IP/port combinations of all related protocols (SMTP, POP3, IMAP) and legacy ports. Over a time span of approx. three months we conducted more than 10 billion TLS handshakes. Additionally, we show that securing server-to-server communication using e.g. SMTP is inherently more difficult than securing client-to-server communication. Lastly, we analyze the volatility of TLS certificates and trust anchors in the e-mail ecosystem and argue that while the overall trend points in the right direction, there are still many steps needed towards secure e-mail.
\end{abstract}

\section{Introduction}
\label{section:introduction}
E-mail has become one of the most fundamental and important services of the Internet and is used by more than a billion users every day. While the implementations and the underlying protocols have (for the most part) endured the test of time, there is no way for users to assess the security and confidentiality of e-mails in transit. Even if the client-to-server connection is secured, server-to-server communication relies on plaintext communication as fallback leaving a large fraction of transmitted e-mails unencrypted and not authenticated. This is a serious limitation, as they are passively observable along the transmission path and users are left in the dark whether their e-mails will be transmitted between servers in plaintext or not. While the advent of opportunistic encryption is good from the privacy viewpoint, it offers no protection against active attackers. Recent data from Google shows that approximately 45\% of inbound and 20\% of outbound e-mail are transmitted in plaintext~\cite{saferEmail} at the time of writing. While PGP would protect e-mails from end to end, its adoption is marginal. Numerous attackers, ranging from local attackers who are able to sniff or modify WiFi traffic, to customer as well as transit ISPs up to nation states are thus able to not only read e-mails in massive quantities, but also to modify, delay or delete them at will. If somehow the usage of TLS between all links during e-mail transit could be enforced, this would not only provide authentication but also confidentiality thanks to the use of encryption. However, due to the decentralized nature of e-mail there is no way to upgrade all servers or somehow enforce link encryption. In fact, we are still just beginning to understand the consequences of large fractions of e-mails being transmitted unencrypted. \\

In this paper we are the first to present a methodology to assess the overall availability of cryptographic primitives for e-mail at scale. While many of the problems of e-mail are connected to the design of popular e-mail protocols, we are the first to quantify and evaluate the adoption of well-known security primitives on a large scale, namely the support of TLS\footnote{In this paper we use the term TLS to refer to all incarnations of the TLS and SSL protocols, if not specified otherwise.} for e-mail protocols on the Internet. Previous large-scale studies on usage and prevalence of TLS focused either on the protocol itself or its usage and implications for secure web browsing using HTTPS. As such, we are the first to present our findings on TLS usage and deployment in e-mail protocols using active probing, with overall more than ten billion TLS handshakes conducted. In particular, the contributions of this paper are as follows: 

\begin{itemize}
    \item We conducted IPv4-wide scans on a number of important ports for e-mail regarding TLS, providing a holistic measurement on e-mail transport layer security.
    \item We compare the results with previous work on HTTPS, and find that TLS in e-mail is inherently less secure.
    \item We analyze the quality and distribution of supported ciphers, arguing that it's still a long way to secure e-mail transport.
    \item We model the graph of all TLS-enabled SMTP servers and find that it is much more complicated to eradicate plaintext and insecure cipher suites then with other protocols.
    \item We evaluate openly available Internet-wide certificate information for TLS in e-mail over a 6-month period.
\end{itemize}

The remainder of this paper is organized as follows: Section~\ref{section:background} gives a brief introduction to TLS and how it is used in e-mail protocols. In Section~\ref{section:scanTLS} we describe our scanning methodology as well as the different inputs we used for seeding our scanning activity. Section~\ref{section:evaluation} presents our results of multiple months of scanning activity, while Section~\ref{section:cipherIslands} is on the problem of insecure ciphers and plaintext in SMTP. Section~\ref{section:discussion} discusses our findings, interprets the results and compares them with other protocols that use TLS. Section~\ref{section:mitigation} discusses mitigation strategies and methods how the overall security for e-mail can be increased. Section~\ref{section:relatedWork} discusses related work, before we conclude in Section~\ref{section:conclusion}.

\section{Background}
\label{section:background}
Today Transport Layer Security (TLS) serves as the underlying foundation for securing many of the most widely used protocols on the Internet, including HTTP and e-mail. TLS provides confidentiality, integrity and authenticity of data transferred on the wire. A general overview of TLS, and more specifically on its usage in HTTPS are given in~\cite{ristic2014bulletproof, clark2013sok}. In recent years the TLS protocol and specific software implementations of the protocol have been under close scrutiny by security researchers, and numerous attacks have been discovered with names like BEAST~\cite{duong2011here}, POODLE~\cite{moller2014poodle}, CRIME~\cite{crime2012}, BREACH~\cite{breach2013} or Lucky13~\cite{al2013lucky} which attack the incorrect usage of block cipher modes, compression and shortcomings in specific versions of TLS. RFC 7457~\cite{rfc7457} provides a good overview of known attacks against TLS. Heartbleed~\cite{durumeric2014heartbleed} on the other hand was a devastating bug in OpenSSL which allowed attackers to read content from the server process' heap memory, including login credentials and the private key used by the server. MS14-066~\cite{MS14-066} (dubbed ``Winshock'') allowed a remote attacker code execution on a system by supplying a malicious ECDSA certificate during the handshake. Other papers analyzed, at large, the unsound use of cryptographic primitives like the Diffie-Hellman key exchange parameters~\cite{adrian2015imperfect} or problematic issues with RC4~\cite{alfardan2013security} or RSA in TLS~\cite{yilek2009private, heninger2012mining, lenstra2012ron}. Lastly it was shown that specific implementations were vulnerable to numerous timing-based side channel attacks~\cite{meyer2014revisiting}. Other publications had a special focus on the usage of TLS on smartphone apps as well as other non-browser software~\cite{fahl2012eve, georgiev2012most, fahl2013rethinking,he2015vetting}. Formal verification of TLS is another research area which received an increased momentum~\cite{avalle2014formal, beurdouche2015messy} as well as the verified implementation of the TLS protocol~\cite{brubaker2014using,bhargavan2013implementing, kaloper2015not}. \\

\begin{figure}[]
	\centering
	\includegraphics[scale=0.60]{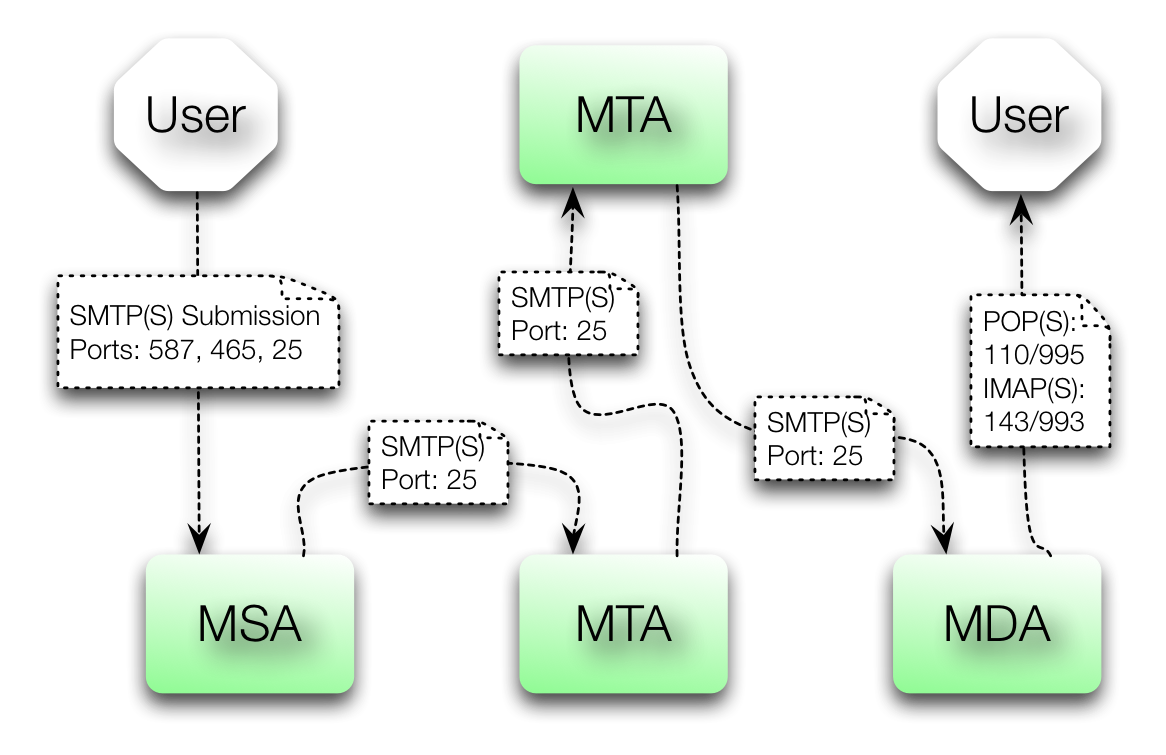}
	\caption{Typical e-mail flow on the Internet from submission to delivery via multiple hops. \emph{MSA: Mail Submission Agent, MTA: Mail Transfer Agent, MDA: Mail Delivery Agent.}}
	\label{fig:mail-flow}
\end{figure}

The e-mail ecosystem makes use of numerous ports and protocols for transmission: SMTP is used for transmitting e-mails between servers, and runs usually over TCP port 25~\cite{rfc5321, rfc3207}. Clients use either the IMAP4 or POP3 protocols to retrieve e-mails from a message delivery server, and SMTP for e-mail submission. IMAP uses TCP ports 143 or 993~\cite{rfc1939, rfc2595}, POP3 uses TCP ports 110 or 995~\cite{rfc3501, rfc2595}, and SMTP Message Submission runs on TCP port 25, 587~\cite{rfc4409} or 465. Although IANA revoked the usage of port 465 for SMTPS in 1998, it is still heavily used for this purpose.  All client-facing protocols are available on two different TCP ports because there are two distinct ways to establish a TLS session: implicit TLS where the connection is established with a TLS handshake, and in-band upgrade of the used plaintext protocol via STARTTLS. Table~\ref{tbl:emailPorts} shows an overview of the commonly used ports for e-mails, including which primitive is used for securing the transport. The complete path an e-mail takes is shown in Figure~\ref{fig:mail-flow}: submitted by the client to the mail submission agent (MSA), the e-mail passes an arbitrary number of relaying mail transfer agent (MTA) servers before it finally reaches the recipient's MTA. Note that the user is only able use or enforce TLS on the edge of this path - she has no control or knowledge whether the MSA or mail delivery agent (MDA) communicate using TLS, or if the MTAs in between make use of it.

\begin{table}
    \begin{center}
    \begin{tabular}{llll}
        \hline
        Port \hspace{0.05cm} & TLS  \hspace{0.05cm} & Protocol  \hspace{0.05cm} & Usage \\
        \hline
        25 & STARTTLS & SMTP & E-mail transmission \\
        110 & STARTTLS & POP3 & E-mail retrieval \\
        143 & STARTTLS & IMAP & E-mail retrieval \\
        465 & implicit & SMTPS & E-mail submission\\
        587 & STARTTLS & SMTP & E-mail submission \\
        993 & implicit & IMAPS & E-mail retrieval \\
        995 & implicit & POP3S & E-mail retrieval \\
		\hline \\
    \end{tabular}
    \caption{Common ports used for e-mail}
    \label{tbl:emailPorts}
    \end{center}
\end{table}

\subsection{TLS handshake and STARTTLS}
\begin{figure}[ht]
	\centering
	\includegraphics[scale=.55]{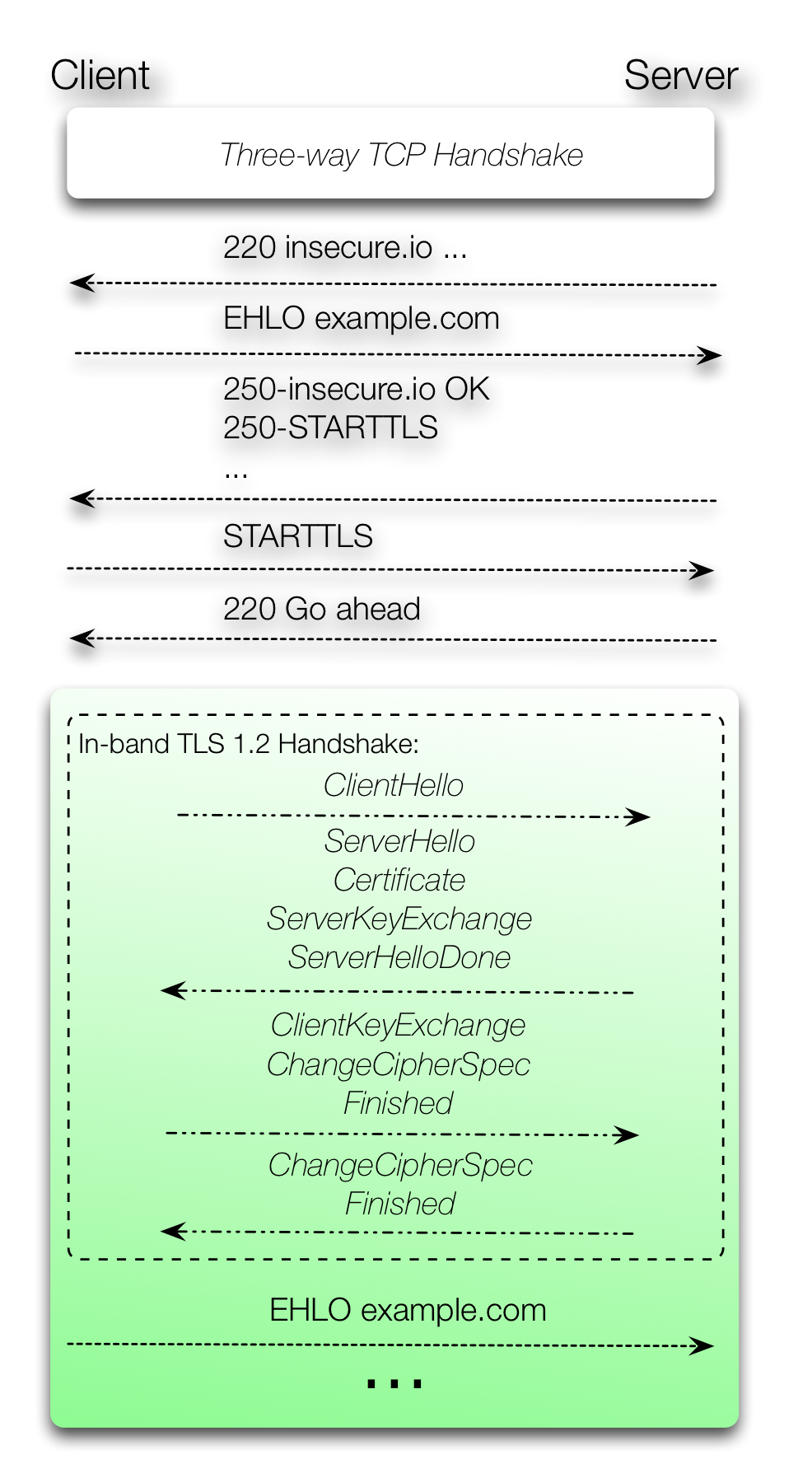}
	\caption{Example of establishing a secured SMTP connection with STARTTLS}
	\label{fig:starttls}
\end{figure}

To initiate a TLS encrypted connection, a TLS handshake is necessary. A ``PreMasterSecret'' is established with a \texttt{ClientKeyExchange} message which is then encrypted using the public key of the server certificate. Client and Server use the ``PreMasterSecret'' together with random numbers (usually a timestamp and random bytes) to establish a common ``MasterSecret''. During this handshake a specific cipher suite is negotiated with \texttt{ChangeCipherSpec} messages. 
A cipher suite is represented by a two byte value and determines the cryptographic primitives to be used in the connection~\cite{rfc2246, rfc3268}. It consists of methods for the key exchange, encryption and message authentication. For example \texttt{TLS\_DHE\_RSA\_WITH\_3DES\_EDE\_CBC\_SHA} (IANA assigned values: \texttt{0x00, 0x16}) is the cipher suite combining an ephemeral Diffie-Hellman key exchange with RSA authentication, 3DES in EDE-CBC mode for encryption and SHA-1 for message authentication (HMAC).
Historically, TLS implementations used their own nomenclature for cipher suites as IANA only standardized TLS parameters in 2006\footnote{https://www.iana.org/assignments/tls-parameters/tls-parameters.xhtml} with the introduction of TLS 1.1.
Figure~\ref{fig:starttls} illustrates a standard in-band TLS handshake (STARTTLS) in the context of SMTP.

The flow of STARTTLS messages is specific to the underlying protocol to be secured and not a general way for in-band upgrade. Different RFCs~\cite{rfc3207,rfc2595} specify these details.

The term ``opportunistic encryption''~\cite{rfc7435} is commonly used when referring to TLS in e-mail protocols. Opportunistic encryption means that connections are usually not 
authenticated (certificates not validated) and are vulnerable to an active attacker but enable more wide-spread use and deployment of security protocols like TLS and
efficiently secure against passive adversaries.

\section{Methodology}
\label{section:scanTLS}
In this section, we describe the source of our input data, the methodology and tools we used, and our considerations regarding the impact of our scans. In the beginning we started with the following requirements: the performance of our methodology should be sufficient in that it runs on commodity hardware with a fast Internet connection. Our goal was to scan more than a million hosts in less than a week, with low additional load for the target systems. Handling network connections is what these systems do (by design), and we took specific precautions that our connections would have no negative impact on the performance or operations. Furthermore our scans should be complete, as in all cipher suites should be covered as well as all the important TLS parameters should be collected i.e. certificates, the certificate chain up to the root CA which signed the certificate, RSA primes and DH-group. For our evaluation we wanted to cover all seven ports which are used by default for e-mail.

\subsection{Input data-set}
The source of our scan consists of a list of all input IP addresses together with the application-specific TCP port. Of course not all IPv4 addresses support an application or even accept incoming data on an application-specific port. We therefore would have to conduct a regular port scan. With the introduction of zmap~\cite{durumeric2013zmap} and similar tools, IPv4-wide port scans can be accomplished quite fast. Many results of port scans are publicly available on a short-term basis, so we decided to use existing datasets from scans.io\footnote{\url{https://scans.io}} as a source for our cipher suite scans. The "Full IPv4 Banner Grab and StartTLS" and the "Full IPv4 TLS Handshake and Banner Grab" datasets from the University of Michigan and the Project Sonar Study (Port 465) from Rapid7 Labs consist of all addressable IPv4 hosts which completed a full TLS handshake at a given time. We used results originating from March to April 2015 as the initial set of IPs for our scans. To further tighten the set of input addresses, we only selected hosts which completed the TLS handshake without any error by filtering the included error key. This resulted in 18,430,143~IP/port combinations which were used for our scan. \\

In June 2015 we verified these IP/port combinations with the results of banner grabbing and certificate collection scans performed with masscan\footnote{\url{https://github.com/robertdavidgraham/masscan}} by our own team. This tool is using asynchronous connections, its own custom user-land TCP/IP stack and other approaches to speed up IP-address and port enumeration. It can also establish a TCP connection to a target system afterwards for e.g. banner collection. This scan resulted in 23,367,809 possible IP/port combinations. Nearly 70\% of these combinations were already included in our initial input data. Because no existing scan for SMTP Port 587 (client-facing message submission) was publicly available, we used the results of a masscan run for input to an additional cipher suite scan on port 587 in June 2015. To allow observations based on the domain of a host, we also scanned MX records for all domains listed in the Alexa Top 1 Million ranking. We stored these records together with the resolved IP addresses to allow the interpretation of our scan result with the popularity of a domain, and to combine them with the Google Transparency Report on safer e-mail~\cite{saferEmail}.

\subsection{Scanning the Cipher Suite Support}
The specification of the TLS handshake protocol~\cite{rfc5246} requires that the client specifies a set of supported cipher suites in the \texttt{ClientHello} handshake message and the server then picks a specific one of them. So the client has to initiate more than one handshake to determine which set of cipher suites for a specific TLS version a server exactly accepts or rejects. The simplest algorithm is to test each cipher suite individually by offering exactly one cipher suite per \texttt{ClientHello}. The response is either a \texttt{ServerHello} or \texttt{Alert} message which can be interpreted as an accepted or rejected cipher suite. We used the tool \emph{sslyze}\footnote{\url{https://github.com/nabla-c0d3/sslyze}} to perform this type of scan. It is a plugin based tool written in Python, built around a custom OpenSSL wrapper. The plugin ``PluginOpenSSLCipherSuites'' was used to test a specific set of cipher suites. Seven different cipher suites were tested for SSLv2 and 136 cipher suites for SSLv3, TLSv1, TLSv1.1 and TLSv1.2 each. This deliberately also includes combinations of TLS version and cipher suite which are not specified in the corresponding RFCs and thus violate the standard e.g. SSLv3 with ECDHE cipher suites. In total we established at least 551 connections per IP to fully test the supported cipher suites for each IP/port combination. We stored the acceptance status as either accepted, rejected, preferred or error. If the handshake was successful we also stored key exchange parameters for the Diffie-Hellman key exchanges. In the case that the TLS connections were not successful, we stored the alert message or error reason given. Additionally we stored the complete certificate chain per IP. \\

To optimize \emph{sslyze} for our needs we made certain changes. \emph{sslyze} was written as a command line tool. To use it efficiently with millions of possible targets we embedded the tool into a distributed environment and integrated it with a queuing system (AMQP). As such our methodology can be easily used in a distributed environment, with many simultaneous hosts that perform scanning activity working together on an input list of IPs to be scanned. We deliberately did not try to hide our scans, and decided to scan the hosts from one host only. We only used one IP for the scans (Ubuntu Linux 14.04, Intel Core i7, 4x 3.10 GHz, 32GB DDR3 RAM, 100MBit/s Network Bandwidth). We consider our results to be a lower bound for assessing the usage of TLS in e-mail, and the scan time can easily be decreased by adding more hosts or bandwidth. Other hosts were used for postprocessing. \\

\emph{sslyze} scans all cipher suites on a per host/port basis, and thus fullfils our requirements listed above. All cipher suite tests are queued consecutive within one process. It then maximizes parallelization by using 15 threads per process and by using more than one process. This intense scan behavior may be seen as a Denial-of-Service attack and may disrupt normal service behavior because all connection attempts are started in a very short timeframe. To solve this issue, we had to spread the cipher suite tests over time. Also, the availability of hosts may change during a scan, which can increase the chance of incomplete results due to hosts that go offline during that timeframe. We therefore serialized the behavior of the scan, spreading the total scan time to several minutes for one host/port, and increased the total number of scan processes. This yielded a good balance between minimizing the scanning impact on the target systems to be scanned -- as not all connections are opened in direct succession -- while completing the entire list of cipher suites to be scanned in reasonable time. With this methodology we were able to scan between 12,000 and 30,000 IP/port combinations per hour. Our bottleneck was identified to be the number of sockets provided by the operating system, in particular the time until a socket can be reused - spreading the load over multiple machines can easily help to overcome this obstacle. After scanning was completed for each port we transformed the output and analyzed it using Google BigQuery\footnote{\url{https://cloud.google.com/bigquery/}}.

\subsection{Ethical Issues}
Similar to related work on active probing on the Internet, we took specific measures to inform not only about our intentions, but also about our scanning methodology and how people can be excluded from future scans. On our scanning host we served a simple webpage to inform about our scans, both over HTTP and HTTPS. We furthermore created a specific abuse address as single point of contact and included it in the WHOIS abuse database. Lastly we set a reverse DNS entry to identify us as a scanning host and to point to the webpage. We were in direct contact with our upstream ISP's network administrators to answer complaints and inquiries and made sure that all requests were answered in a timely manner. To prevent connection- or resource-wise Denial of Service for the hosts to be scanned, we tested our scanning methodology on devices with constrained hardware, in addition to spreading the number of connections over time. These devices were limited in the overall computation resources. We choose them to reflect, appropriately, the smallest possible devices on the Internet that could be used to run a full e-mail stack, similar to a very small cloud instance (VPS) or an embedded system (router or appliances). We did not take any specific measures to prevent our connections from being logged, and did not take any measures to stay below the radar of vigilant system administrators. \\

Overall, we received 89 e-mails on our abuse e-mail address. 52 of them were automatically generated by IDS tools to inform us that one of our hosts might be engaging in some possibly unsolicited scanning activity and to investigate the issue accordingly. 16 e-mails were simple blacklisting requests, and the other e-mails were mostly driven by curiosity where administrators wanted to know additional details on our scans and in particular how their setup performed compared to the rest. A few e-mails were just blatantly rude. Upon investigating one of the complaints in depth it was discovered that there was a bug in the dovecot mail server that caused a sub-process of the software to crash\footnote{CVE-2015-3420}. This bug was fixed by the software maintainers shortly after notification. 


\section{Evaluation}
\label{section:evaluation}
This section presents the data we obtained through our scans. First we give a general overview on the datasets we obtained. Next we evaluate the cipher suite acceptance and cipher suite preference of the scanned hosts. We then take a closer look at the accepted cryptographic primitives and the parameters used in Diffie-Hellman key exchanges. Lastly we present pressing problems with how TLS is used in related e-mail protocols.


In total we conducted 20,270,768 scans over seven different TCP ports (Table~\ref{tab:dataset-overview}). The scans were performed between April and August 2015. Thereof, 18,381,936 were valid results, meaning that at least one TLS handshake was completed. 1,888,832 of the scans were invalid with no TLS session established at all (77.9\%~Could not connect (timeout), 17.5\%~Connection rejected, 3\%~SMTP~EHLO rejected, 1.6\%~STARTTLS rejected). Since we established 551 TLS handshakes with different cipher suites and TLS versions per host and port, this resulted in more than 10.2 billion TLS handshakes conducted. 89.78\% of these handshakes got rejected, 8.26\% accepted and 1.95\% resulted in an error. The high percentage of rejected ciphers is due to the fact that e-mail servers typically do not support all TLS versions available between SSLv2 and TLS 1.2, and not every cipher suite for each supported TLS version.

\begin{table*}[t]
\begin{center}
\begin{tabular}{llrrrrrr}
\hline
Port   & Protocol    & Source of Input & Timeframe   & Total scans & Valid results & Invalid resuls & Reoccured in masscan results (July)\\
\hline
25     & SMTP with STARTTLS       & scans.io UMich    & 05/05 -- 05/20 & 1,825,657 & 1,558,796 & 266,861 &  1,611,051 \\
110    & POP3 with STARTTLS       & scans.io UMich    & 05/20 -- 06/04 & 4,123,819 & 3,553,201 & 570,618 &  3,572,513 \\
143    & IMAP with STARTTLS       & scans.io UMich    & 06/04 -- 06/06 &   859,470 &   764,211 &  95,259 &    765,511 \\
465    & SMTP over TLS            & scans.io Rapid7 & 06/07 -- 06/21 & 3,123,874 & 2,851,598 & 272,276 &  2,681,955 \\
587    & Submission with STARTTLS & masscan results & 07/27 -- 08/03 & 1,840,629 & 1,656,901 & 183,728 &          - \\
993    & IMAP over TLS            & scans.io UMich    & 04/27 -- 05/05 & 4,254,564 & 4,047,385 & 207,179 &  3,897,348 \\
995    & POP3 over TLS            & scans.io UMich    & 04/17 -- 04/27 & 4,242,755 & 3,949,844 & 292,911 &  3,775,292 \\
\hline
       & $\Sigma$                 &                 &               & 20,270,768 & 18,381,936 & 1,888,832 & \\ 
\hline \\
\end{tabular}
\end{center}
\caption{Overview of our Scan Results}
\label{tab:dataset-overview}
\end{table*}


\subsection{Supported cipher suites}
\label{ssec:ciphers}

Table~\ref{tab:protocol-versions} shows the percentage of unique hosts that support a specific TLS protocol version. A host supports a TLS version if at least one handshake with any chosen cipher suite for this version is accepted by the server. SMTP is remarkable in that it has a relatively large acceptance of SSLv2 and SSLv3. The support of SSLv2 for client-to-server connections is generally low, except for e-mail submission on port 587 where 15\% of all tested IPs accept cipher suites from SSLv2. TLSv1 is with more than 90\% clearly the most supported version for all studied e-mail protocols. This means that 9 out of 10 e-mail servers that speak TLS can use TLSv1, whereas approximately 50--66\% accept the newer versions TLSv1.1 and TLSv1.2. \\

Table~\ref{tab:protocol-versions} shows the percentages of exclusive TLS version subsets. Hosts that support only TLSv1.1 and TLSv1.2 are close to non-existing across all protocols and SMTP on port 25 using exclusively other TLS version combinations as listed in the table. It also shows some TLS version support for combinations of interest e.g, SSLv2 and SSLv3 without any support for TLSv1, TLSv1.1 and TLSv1.2 (which is close to non-existing) or modern configurations without support of SSLv2 or SSLv3 is deployed by up to one third of all hosts for mail delivery protocols. Only 8\% of all SMTP hosts have this configuration.


\begin{table*}[]
\centering
\begin{tabular}{p{7cm}VVVVVVV}
\hline
TLS version & 25     & 110       & 143      & 465       & 587      & 993       & 995       \\
\hline
SSLv2   & 41.7\%  & 2.9\%    & 3.7\%    & 4.2\%     & 15.0\%     & 8.7\%     & 9.3\%     \\
SSLv3   & 82.6\%  & 35.3\%   & 38.6\%   & 36.1\%    & 64.0\%     & 53.7\%    & 53.8\%    \\
TLSv1   & 91.6\%  & 98.9\%   & 98.9\%   & 94.8\%    & 92.4\%     & 98.2\%    & 97.9\%    \\
TLSv1.1 & 55.7\%  & 50.5\%   & 54.1\%   & 65.6\%    & 49.6\%     & 57.2\%    & 55.1\%    \\
TLSv1.2 & 56.1\%  & 50.7\%   & 54.2\%   & 65.2\%    & 48.8\%     & 57.7\%    & 55.6\%    \\
\hline
only SSLv2\&3  & 0.2\%  &  0.0\% &  0.0\% & 0.0\% & 0.0\%      &  0.1\% &  0.1\% \\
only TLSv1 & 1.6\% & 27.2\% &  23.1\% & 13.9\% & 10.6\% & 12.1\% & 12.6\% \\
only TLSv1--1.2 & 8.0\%  & 36.4\% & 37.2\% & 44.7\% & 17.9\%      & 32.8\% & 31.9\% \\
only SSLv3\&TLSv1  & 7.0\%  & 18.4\% & 19.1\% & 11.1\%  & 21.1\%     & 22.3\% & 22.9\% \\
only TLSv1.1--1.2 & 0.1\%  & 0.1\%  & 0.1\%  & 0.0\%  & 0.5\%  & 0.1\%  & 0.1\%  \\
\hline \\
\end{tabular}
\caption{Protocol version support}
\label{tab:protocol-versions}
\end{table*}

In Table~\ref{tab:accepted-cipher suites} we show the acceptance rate of selected cipher suites for TLSv1 (as it is by far the most supported version of TLS across all protocols).

For many cipher suites we can identify a big gap in the acceptance rate of weak cipher suites for SMTP and client protocols. E.g., DES-CBC-SHA is only supported by 6\% of all IMAP and POP hosts, but 75\% of SMTP hosts. The percentage for submission lies at 29\%. 
To better understand these results, we cluster the cryptographic primitives of all accepted cipher suites in the next subsection.

\begin{table*}[]
\centering
\begin{tabular}{p{7cm}VVVVVVV}
\hline
cipher suite name (OpenSSL) & 25      & 110     & 143     & 465     & 587     & 993     & 995     \\
\hline
AES256-SHA           & 89.0\% & 91.5\% & 93.1\% & 92.8\% &  82.7\%  & 89.4\% & 88.4\% \\
CAMELLIA256-SHA      & 57.3\% & 54.4\% & 57.2\% & 65.3\% &  49.5\% & 55.3\% & 53.7\% \\
DES-CBC-SHA          & 75.0\% & 5.9\%  & 5.6\%  & 6.7\%  & 29.3\%  & 6.1\%  & 6.2\%  \\
DES-CBC3-SHA         & 89.4\% & 95.9\% & 97.1\% & 92.6\% & 90.8\% & 93.4\% & 93.1\% \\
DHE-RSA-AES256-SHA   & 83.6\% & 72.0\% & 73.5\% & 85.4\% & 63.5\% & 72.5\% & 71.8\% \\
IDEA-CBC-SHA         & 21.3\% & 45.0\% & 46.4\% & 46.0\% & 28.1\% & 45.3\% & 45.3\% \\
SEED-SHA             & 52.8\% & 47.6\% & 49.8\% & 52.4\% & 35.3\% & 48.3\% & 47.2\% \\
RC4-SHA              & 84.9\% & 85.1\% & 85.4\% & 81.6\% & 78.8\% & 81.7\% & 81.9\% \\
ECDHE-RSA-AES256-SHA & 39.8\% & 8.1\%  & 8.4\%  & 11.3\% & 24.1\%  & 8.6\%  & 7.0\%  \\
ECDHE-RSA-RC4-SHA    & 38.4\% & 6.0\%  & 6.4\%  & 10.4\% & 18.8\%  & 6.5\%  & 5.7\%  \\
EXP-RC4-MD5          & 72.7\% & 17.9\% & 16.4\% & 5.2\%  & 28.0\% & 14.1\% & 14.0\% \\
EXP-RC2-CBC-MD5      & 72.4\% & 17.7\% & 16.2\% & 5.2\%  & 27.7\% & 14.1\% & 14.0\% \\
EXP-DES-CBC-SHA      & 72.7\% & 17.6\% & 16.1\% & 5.2\%  & 27.7\% & 14.0\% & 13.9\% \\
\ldots               & \ldots  & \ldots  & \ldots  & \ldots & \ldots & \ldots & \ldots \\
\hline \\
\end{tabular}
\caption{Percentage of selected cipher suite acceptance for TLSv1.}
\label{tab:accepted-cipher suites}
\end{table*}

To test preferred cipher suites, we specified a set of cipher suites in the \texttt{ClientHello} message. cipher suites with CAMELLIA and 3DES encryption were not considered due to a bug in some already deployed versions of \emph{libssl}\footnote{\url{https://bugs.debian.org/cgi-bin/bugreport.cgi?bug=665452}}. Table~\ref{tab:preferred-cipher suites} shows how often a specific cipher suite was chosen. 44\% of all SMTP deployments prefer the ECDHE-RSA-AES256-SHA cipher suite over DHE-RSA-AES256-SHA. Static RSA key exchange with AES256-SHA is only preferred by 5\%. For all other protocols 60--80\% of all hosts prefer DHE-RSA-AES256-SHA over its counterpart with ECDHE or static RSA.

\begin{table*}[]
\centering
\begin{tabular}{p{1.5cm}p{5cm}VVVVVVV}
\hline
Version & cipher suite                 & 25      & 110     & 143     & 465    & 587      & 993     & 995     \\
\hline
TLSv1    & DHE-RSA-AES256-SHA          & 49.64\% & 68.03\% & 67.89\% & 79.32\%& 47.72\%      & 68.39\% & 69.65\% \\
TLSv1    & ECDHE-RSA-AES256-SHA        & 43.67\% & 6.44\%  & 6.84\%  & 11.49\%& 23.01\%      & 7.43\%  & 6.13\%  \\
TLSv1    & AES256-SHA                  & 4.94\%  & 17.67\% & 17.89\% & 7.17\% & 16.41\%     & 17.23\% & 17.25\% \\
\hline \\
\end{tabular}
\caption{Top preferred cipher suites for TLSv1. (minus 3DES and CAMELLIA)}
\label{tab:preferred-cipher suites}
\end{table*}


\subsection{Cryptographic primitives}
\label{ssec:cryptographic-primitives}

Table~\ref{tab:keyexchange} compares the different percentages of key exchange methods used. Static RSA is supported by the majority of hosts. 15\%-17\% of POP3 and IMAP deployments use static RSA as the only key exchange algorithm. Just a few (7--9\%) of all POP3 and IMAP hosts offer ECDHE for key exchange, but close to 75\% support DHE as key exchange mechanism. Both methods guarantee perfect forward secrecy. A high number of deployments accept weak anonymous Diffie-Hellman and export grade handshakes.

\begin{table*}[h]
\centering
\begin{tabular}{p{7cm}VVVVVVV}
\hline
            & 25     & 110    & 143    & 465    & 587    & 993    & 995    \\
\hline
RSA         & 92.3\% & 99.1\% & 99.0\% & 95.0\% & 93.0\%  & 99.1\% & 99.0\% \\
DHE-RSA     & 85.9\% & 74.8\% & 74.7\% & 86.4\% & 65.0\%  & 73.9\% & 73.7\% \\
ECDHE-RSA   & 41.2\% & 8.4\%  & 8.7\%  & 11.5\% & 25.2\%  & 8.9\%  & 7.2\%  \\
ADH         & 64.3\% & 7.8\%  & 6.3\%  & 14.5\% & 26.0\%  & 5.3\%  & 5.5\%  \\
AECDH       & 35.9\% & 1.7\%  & 1.7\%  & 7.3\%  & 14.5\%  & 1.7\%  & 1.6\%  \\
EXP-RSA     & 75.2\% & 18.1\% & 16.5\% & 5.8\%  & 28.9\%  & 15.7\% & 16.1\% \\
EXP-DHE-RSA & 71.6\% & 11.8\% & 10.3\% & 3.6\%  & 18.7\%  & 9.2\%  & 9.4\%  \\
EXP-ADH     & 63.2\% & 7.5\%  & 6.0\%  & 0.1\%  & 12.8\%  & 5.0\%  & 5.2\%  \\
\hline
Only RSA   & 2.7\%  & 15.9\%  & 16.2\%  & 6.2\%  & 14.5\%  & 17.1\%  & 17.4\%  \\
\hline \\
\end{tabular}
\caption{Key exchange method support}
\label{tab:keyexchange}
\end{table*}

Table~\ref{tab:cipher} presents the percentage of hosts that support an encryption method with given key size. 
3DES and AES in CBC mode are currently the most widely adopted encryption methods.
Because of several weaknesses~\cite{alfardan2013security} the use of RC4 in TLS was prohibited in February 2015~\cite{rfc7465}. We observe that it is still widely supported by 82.3\% to 86.5\% of all hosts. However, less than 0.9\% exclusively support RC4.
AES in GCM is supported by 49.2\% to 63.3\% hosts. Still 15--18\% of all hosts support export ciphers (DES-40, RC4-40, RC2-40) for POP3 and IMAP, 73--75\% for STMP and 28--30\% for submission.

\begin{table*}[h]
\centering
\begin{tabular}{p{7cm}VVVVVVV}
\hline
Encryption algorithm    & 25     & 110    & 143    & 465    & 587     & 993    & 995    \\
\hline
AES-128 (CBC) & 91.3\% & 93.5\% & 94.0\% & 93.5\% & 84.3\%     & 92.2\% & 91.3\% \\
AES-256 (CBC) & 91.6\% & 93.4\% & 93.8\% & 93.9\% & 84.5\%     & 93.4\% & 92.5\% \\
3DES-168      & 91.6\% & 97.4\% & 97.5\% & 93.3\% & 92.3\%     & 96.2\% & 96.3\% \\
RC4-128       & 86.3\% & 86.5\% & 85.7\% & 82.3\% & 79.6\%     & 83.4\% & 83.7\% \\
CAMELLIA-256  & 58.8\% & 55.9\% & 57.8\% & 66.0\% & 51.3\%     & 58.2\% & 56.5\% \\
CAMELLIA-128  & 58.7\% & 55.9\% & 57.8\% & 66.0\% & 51.3\%     & 58.5\% & 56.9\% \\
AES-128 (GCM) & 54.9\% & 49.3\% & 52.8\% & 63.3\% & 46.4\%     & 55.2\% & 53.4\% \\
AES-256 (GCM) & 55.4\% & 49.2\% & 52.7\% & 63.0\% & 46.6\%     & 55.1\% & 53.0\% \\
SEED-128      & 53.8\% & 48.6\% & 50.0\% & 52.8\% & 35.8\%     & 49.1\% & 47.9\% \\
IDEA-128      & 21.6\% & 45.9\% & 46.5\% & 46.3\% & 28.4\%     & 45.8\% & 45.9\% \\
DES-56        & 76.1\%  & 5.8\%  & 5.6\%  & 6.8\%  & 29.8\%     & 6.5\%  & 6.8\%  \\
RC4-40        & 74.9\%  & 18.1\% & 16.5\% & 5.8\%  & 28.8\%     & 15.4\% & 15.7\% \\
RC2-40        & 74.9\%  & 17.9\% & 16.3\% & 5.8\%  & 28.7\%     & 15.3\% & 15.6\% \\
DES-40        & 73.6\%  & 17.7\% & 16.2\% & 5.2\%  & 28.1\%     & 14.7\% & 14.8\% \\
RC2-128       & 40.9\%  & 2.7\%  & 2.2\%  & 4.1\%  & 14.4\%     & 5.6\%  & 6.3\%  \\
NULL          & 0.1\% & 0.00\% & 0.01\% & 0.02\% & 0.02\%     & 0.05\% & 0.04\% \\
\hline
Only AES-CBC  & 0.16\% & 0.11\% & 0.07\% & 0.06\% & 0.11\% & 0.22\% & 0.37\% \\
Only RC4      & 0.05\% & 0.45\% & 0.45\% & 0.87\% & 0.10\% & 0.78\% & 0.80\%  \\
\hline \\
\end{tabular}
\caption{Encryption method support}
\label{tab:cipher}
\end{table*}

\subsection{Key Exchange Parameters}
\label{ssec:handshakes}

The recently discovered Logjam attack against TLS~\cite{adrian2015imperfect} was possible because many servers employ weak Diffie-Hellman parameters. We analyzed used group sizes, since 512 bit and 768 bit groups are considered very weak. Common used primes are more valuable for precalculation attacks, so we tried to uncover some of them. 

\subsubsection{Weak Diffie-Hellman parameters}
We analyzed the use of different Diffie-Hellman parameters and thereafter categorized all cipher suites into two sets: cipher suites with export restricted algorithms and all other cipher suites.
Because the Diffie-Hellman key exchange for export cipher suites is restricted to 512 bit, almost all use this size. For SMTP 12 hosts used a different group size, whereas 1,045,456 hosts used the 512 bit group.
For non-export cipher suites, 1024 bit parameters are clearly most commonly used. Still 6-7\% of POP3 and IMAP use 768 bit parameters. These parameters will soon be deprecated in OpenSSL.\footnote{\url{https://www.openssl.org/blog/blog/2015/05/20/logjam-freak-upcoming-changes/}} 2048 bit parameters are rarely deployed and 4096 are close to non-existing. One exception are parameters used on TCP port 465. 69.5\% of all hosts use a 2048 bit group and only 30.39\% use a 1024 bit group. 
We explain this difference with the default configuration of Exim, a popular MTA daemon which uses a 2048 bit DH group (\emph{2048-bit MODP Group with 224-bit Prime Order Subgroup}~\cite{rfc5114}) by default.

\begin{table*}[]
\centering
\begin{tabular}{p{5cm}p{1.5cm}VVVVVVV}
\hline
cipher suites & Size of Prime   & 25       & 110      & 143      & 465      & 587     & 993      & 995      \\
\hline
Export & 512      & 100\%    & 100\%   & 100\% &   100\% & 100\%  & 100\% & 100\% \\
Non-Export & 512  & 0.08\%   & 0.04\%   & 0.05\%   & 0.05\%   &  0.05\%  & 0.07\%   & 0.05\%   \\
Non-Export & 768  & 0.02\%   & 6.31\%   & 6.52\%   & 0.00\%   &  0.03\%  & 7.02\%   & 7.43\%   \\
Non-Export & 1024 & 99.10\%  & 92.11\%  & 91.64\%  & 30.39\%  & 75.39\%  & 92.14\%  & 91.77\%  \\
Non-Export & 2048 & 0.79\%   & 1.50\%   & 1.75\%   & 69.50\%  & 24.48\%  & 0.71\%   & 0.70\%   \\
Non-Export & 4096 & 0.01\%   & 0.00\%   & 0.01\%   & 0.01\%   &  0.04\%  & 0.01\%   & 0.00\%   \\
\hline \\
\end{tabular}
\caption{DH Parameter Size}
\label{tab:dh-param}
\end{table*}

\subsubsection{Reuse of common primes for Diffie-Hellman}

For SMTP we found one 512 bit prime which is used by 996,792 distinct IPs (64\% of all 1,557,288 valid scans) and also one 1024 bit prime which is used by 1,077,736 distinct IPs (69.2\%). These two primes are statically included in the postfix source code.\footnote{\url{https://github.com/vdukhovni/postfix/blob/master/postfix/src/tls/tls\_dh.c}}
We also rediscovered these primes (512 bit: 2,066 hosts, 1024 bit: 512,391 hosts) at DH handshakes on port 465.
1,673,271 distinct IPs on port 465 (58.7\%) make use of the specific DH group used by Exim\footnote{\url{https://github.com/Exim/exim/blob/master/src/src/std-crypto.c}} as described above.
On port 587, 539,322 IPs (32.7\%) use the 1024 bit and 212,898 IPs (12.9\%) use the 512 bit prime used by postfix. 218,163 (13.2\%) use the Exim 2048 bit group and 85,548 (5.2\%) use a 1024 bit prime included in the nginx source code\footnote{\url{http://trac.nginx.org/nginx/browser/nginx/src/event/ngx\_event\_openssl.c}}.
99,278 IPs on port 993 (2.38\%) and 94,555 IPs on port 995 (2.4\%) use the same 1024 bit nginx prime. 512 bit primes are not shared for these protocols. Only 721 IPs (IMAPS) and 733 IPs (POP3S) share the most used 512 bit prime.
Primes used in server deployments for POP and IMAP are more diverse. The most used 1024 bit prime is used by 27,429 IPs (POP) and 6,291 IPs (IMAP), which are only 0.8\% of all distinct IPs each. The most commonly shared 512 bit prime is only used by 0.05\% of all IPs.

\subsubsection{Elliptic Curve Diffie-Hellman}

One alternative is using Diffie-Hellman key exchanges based on elliptic curve cryptography.
We examined common used curves specified for the use in TLS~\cite{rfc4492}.
For SMTP related protocols the secp256r1 curve is used for up to 99.2\% of all ECDH handshakes. For POP3 and IMAP ports the usage differs, as 3 different curves are in use whereas secp384r1 is by far the most used curve. Additionally, the curves sect163r2, secp224r1 and secp256k1 occur, but only for less than 0.03\% of all handshakes.

\begin{table*}[]
\centering
\begin{tabular}{p{7cm}VVVVVVV}
\hline
DH Curve   & 25      & 110     & 143     & 465     & 587     & 993     & 995     \\
\hline
secp256r1 & 99.20\% & 27.67\% & 25.52\% & 99.73\% & 94.63\% & 28.77\% & 23.26\% \\
secp384r1 & 0.22\%  & 65.70\% & 67.35\% & 0.10\%  & 3.84\%  & 63.82\% & 69.65\% \\
secp521r1 & 0.12\%  & 0.94\%  & 1.01\%  & 0.12\%  & 0.93\%  & 1.02\%  & 0.93\%  \\
sect571r1 & 0.46\%  & 5.69\%  & 6.13\%  & 0.02\%  & 0.59\%  & 6.38\%  & 6.13\%  \\
\hline \\
\end{tabular}
\caption{Percentage of different Elliptic Curves used}
\label{tab:ecdh-curves}
\end{table*}


\subsection{TLS Certificates}

The subject of certificate validation for HTTPS has been investigated by various other studies~\cite{holz2011ssl,eckersley2010observatory,durumeric2013analysis}.
In our study we analyzed 2,115,228 different unique leaf certificates for all e-mail related protocols. We used certificates collected with our cipher suite scans. For temporal analysis we additionally used scans conducted by University of Michigan. 


\subsubsection{Self-signed certificates}

We used the Mozilla NSS Truststore (as of 04/2015) to validate all certificates. 
For client-to-server protocols the percentage of self-signed certificates used is very high. Over 50\% of all distinct IPs use self-signed certificates, only 33--37\% are correctly validated. For SMTP the number of self-signed certificates is even higher. 65\% of all distinct IPs we scanned use a self-signed certificate (expired self-signed certificates included). This is no surprise, since SMTP deployments do not validate certificates by default. Around 1\% of all valid certificates are valid but expired. 
An overview is given in Table~\ref{tab:cert-truststore}. 

\begin{table*}[]
\centering
\begin{tabular}{p{7cm}VVVVVVV}
\hline
Result of Validation      & 25     & 110     & 143     & 465     & 587     & 993     & 995     \\
\hline
ok                                           & 20.66\% & 32.87\% & 34.11\% & 36.20\% & 37.57\% & 35.81\% & 37.43\% \\
self-signed certificate                      & 65.20\% & 52.13\% & 53.05\% & 53.96\% & 38.61\% & 51.57\% & 50.03\% \\
unable to get local issuer certificate       & 10.05\% & 11.31\% & 11.01\% & 8.30\%  & 20.50\% & 11.09\% & 11.19\% \\
validation error                             & 1.24\%  & 2.49\%  & 0.53\%  & 0.24\%  & 0.99\%  & 0.07\%  & 0.08\%  \\
certificate has expired                      & 0.98\%  & 0.77\%  & 0.81\%  & 0.75\%  & 1.26\%  & 0.86\%  & 0.80\%  \\
\hline \\
\end{tabular}
\caption{Truststore results for all distinct IPs in Mozilla NSS Truststore}
\label{tab:cert-truststore}
\end{table*}

\subsubsection{RSA public key size}

We find that over 99\% of all leaf certificates use RSA public keys. 
We analyzed the public key size of these certificates and find that over 90\% of all trusted leaf certificates are equipped with 2048 bit public keys. Less than 10\% use 4096 bit public keys and less than 0.1\% use 1024 bit keys. This situation changes if we consider self-signed certificates. 1024 bit keys are often used (15--40\%), followed by 2048 bit keys. 

\begin{table*}[]
\centering
\begin{tabular}{p{7cm}VVVVVVV}
\hline
RSA Public Key Size            & 25      & 110     & 143     & 465     & 587     & 993     & 995     \\
\hline
\multicolumn{7}{l}{All Certificates} \\
\hline
512                            & 0.29\%  & 0.13\%  & 0.13\%  & 0.06\%  & 0.10\%     & 0.10\%  & 0.05\%  \\
1024                           & 19.97\% & 17.57\% & 16.37\% & 9.85\%  & 19.01\%    & 16.27\% & 15.29\% \\
2048                           & 75.74\% & 79.49\% & 80.45\% & 86.68\% & 76.75\%    & 80.28\% & 81.58\% \\
4096                           & 3.78\%  & 2.72\%  & 2.96\%  & 3.34\%  & 4.02\%     & 3.27\%  & 3.00\%  \\
\hline
\multicolumn{7}{l}{Trusted Leaf Certificates (OK)} \\
\hline
1024                           & 0.06\%  & 0.03\%  & 0.02\%  & 0.03\%  & 0.04\%     & 0.04\%  & 0.05\%  \\
2048                           & 92.59\% & 94.61\% & 94.34\% & 93.54\% & 93.73\%    & 93.93\% & 94.35\% \\
4096                           & 7.24\%  & 5.34\%  & 5.60\%  & 6.40\%  & 6.18\%     & 5.99\%  & 5.58\%  \\
\hline
\multicolumn{7}{l}{Self Signed Certificates} \\
\hline
512                            & 0.17\%  & 0.03\%  & 0.03\%  & 0.04\%  & 0.05\%     & 0.05\%  & 0.04\%  \\
1024                           & 23.14\% & 30.28\% & 28.26\% & 14.06\% & 39.32\%    & 28.26\% & 27.62\% \\
2048                           & 74.17\% & 68.45\% & 70.38\% & 84.67\% & 57.82\%    & 70.30\% & 71.10\% \\
4096                           & 2.26\%  & 1.11\%  & 1.19\%  & 1.14\%  & 2.58\%     & 1.29\%  & 1.12\%  \\
\hline \\
\end{tabular}
\caption{RSA Public Key Size}
\label{tab:rsa-pks}
\end{table*}

\subsubsection{Common leaf certificates}

Many certificates are found on more than one IP. Table~\ref{tab:common-certificates} shows the number of IPs that use the same certificate. The table includes the top certificates per port. We identify certificates that are used by a massive number of IPs, usually in same subnets. The certificate with the SHA1 fingerprint \texttt{b16c\ldots6e24} was provided on 85,635 IPs in 2 different /16 IP ranges and on 2 IPs in two completely different IP ranges. 

\begin{table}[]
\centering
\begin{tabular}{lllr}
\hline
Common Name (Issuer Common Name) & Fingerprint & Port & IPs \\
\hline
*.nazwa.pl (nazwaSSL) 							& b16c\ldots6e24 & 25 &  40,568 \\
                                                &                & 465 & 81,514 \\
                                                &                & 587 & 84,318 \\
                                                &                & 993 & 85,637 \\
                                                &                & 995 & 85,451 \\
\hline
*.pair.com (USERTrust RSA Organization \ldots)  & a42d\ldots768f & 25 & 15,573  \\
                                               	&                & 110 & 60,588 \\
                               				    &                & 143 & 13,186 \\
                                                &                & 465 & 63,248 \\
                                                &                & 587 & 61,933 \\
                                                &                & 993 & 64,682 \\
                                                &                & 995 & 64,763 \\
\hline
*.home.pl (RapidSSL SHA256 CA - G3)             & 8a4f\ldots6932 & 110 & 126,174 \\
                                                &                & 143 & 26,735  \\
                                                &                & 587 & 125,075 \\
*.home.pl (AlphaSSL CA - SHA256 - G2)           & c4db{\ldots}a488 & 993 & 128,839  \\
                                                &                  & 995 & 126,102  \\
\hline
*.sakura.ne.jp (RapidSSL SHA256 CA - G3)        & 964b{\ldots}c39e & 25 & 16,573 \\
\hline 
*.prod.phx3.secureserver.net (Starfield \ldots)	& f336{\ldots}ac57 & 993 & 61,307 \\
                                                &                  & 995 & 61,250 \\
\hline \\ 
\end{tabular}
\caption{Common leaf certificates}
\label{tab:common-certificates}
\end{table}

We also found self-signed certificates that do not share the same fingerprint, but subject attributes are the same. (\texttt{CommonName, OrganizationalUnitName, OrganizationalName}). We identify entities that are responsible for a large number of self-signed certificates. These entities are either software administration tools like \emph{Parallels Panel} or \emph{Plesk} or default configurations of mail software like \emph{Courier} and \emph{Dovecot}. They also play an important role in reducing the use of self-signed certificates in the future.
An overview of top entitities is shown in Table~\ref{tab:ssc-common-issuer}.

\begin{table}[]
\centering
\begin{tabular}{lrrrr}
\hline
Name      & Key Size & IPs \\
\hline
Parallels Panel - Parallels                                & 2048        & 306,852  \\
imap.example.com - IMAP server                             & 1024        & 261,741  \\
Automatic{\ldots}POP3 SSL key - Courier Mail Server        & 1024        &  87,246   \\
Automatic{\ldots}IMAP SSL key - Courier Mail Server        & 1024        &  83,976   \\
Plesk - Parallels                                          & 2048        &  68,930   \\
localhost.localdomain - SomeOrganizationalUnit             & 1024        &  26,248   \\
localhost - Dovecot mail server                            & 2048        &  13,134   \\
plesk - Plesk - SWsoft, Inc.                               & 2048        &  14,207   \\
\hline \\      
\end{tabular}
\caption{Top self signed certificiates for all protocols}
\label{tab:ssc-common-issuer}
\end{table}

\subsubsection{Weak RSA Keys}
\label{ssec:certificates}
Of 30,757,242 collected RSA certificates from all e-mail protocols and related ports we've extracted
2,354,090 unique RSA moduli and were able to recover 456 GCDs (hence RSA private keys). We used techniques and
tooling provided by the authors of~\cite{heninger2012mining} for computation of vulnerable moduli. A similar
analysis was performed on data-sets for HTTPS in \cite{durumeric2013analysis}.


\subsection{Certificate volatility}
\label{ssec:volatility}

\begin{figure}[]
	\centering
	\includegraphics[width=\columnwidth]{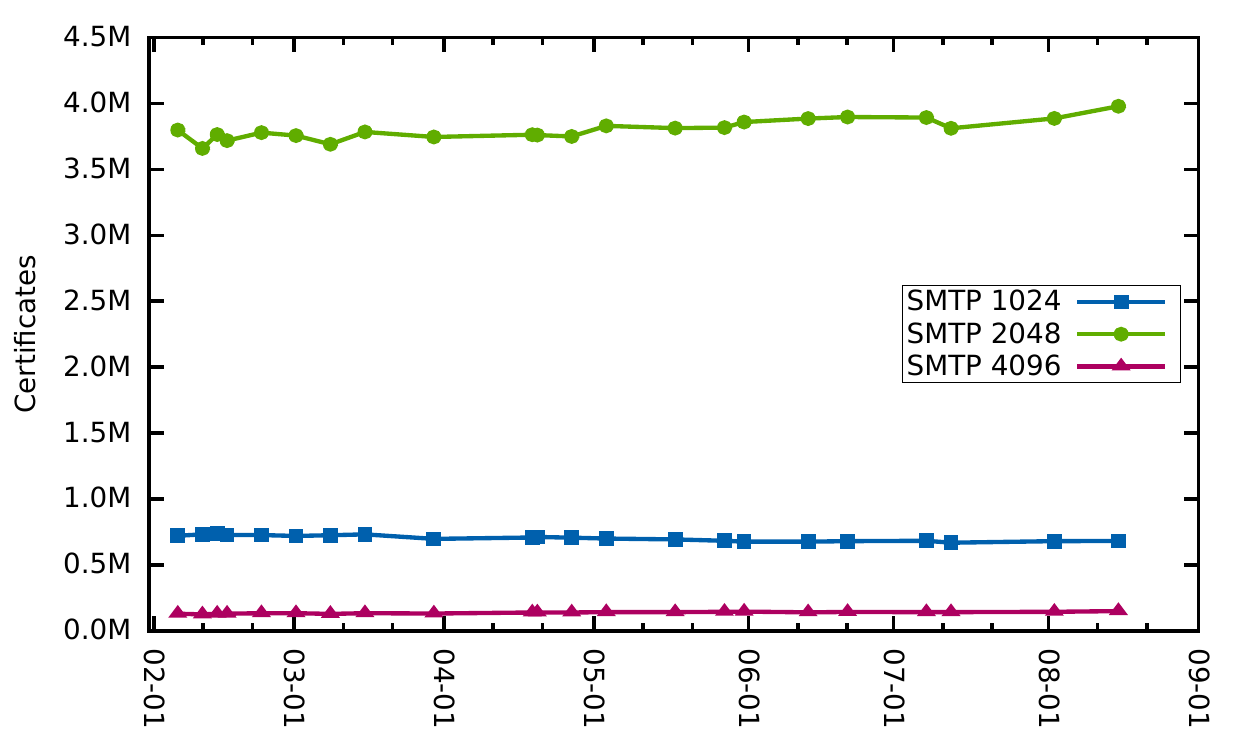}
	\caption{SMTP leaf certificate volatility}
	\label{fig:smtp_volatility}
\end{figure}

\begin{figure}[]
	\centering
	\includegraphics[width=\columnwidth]{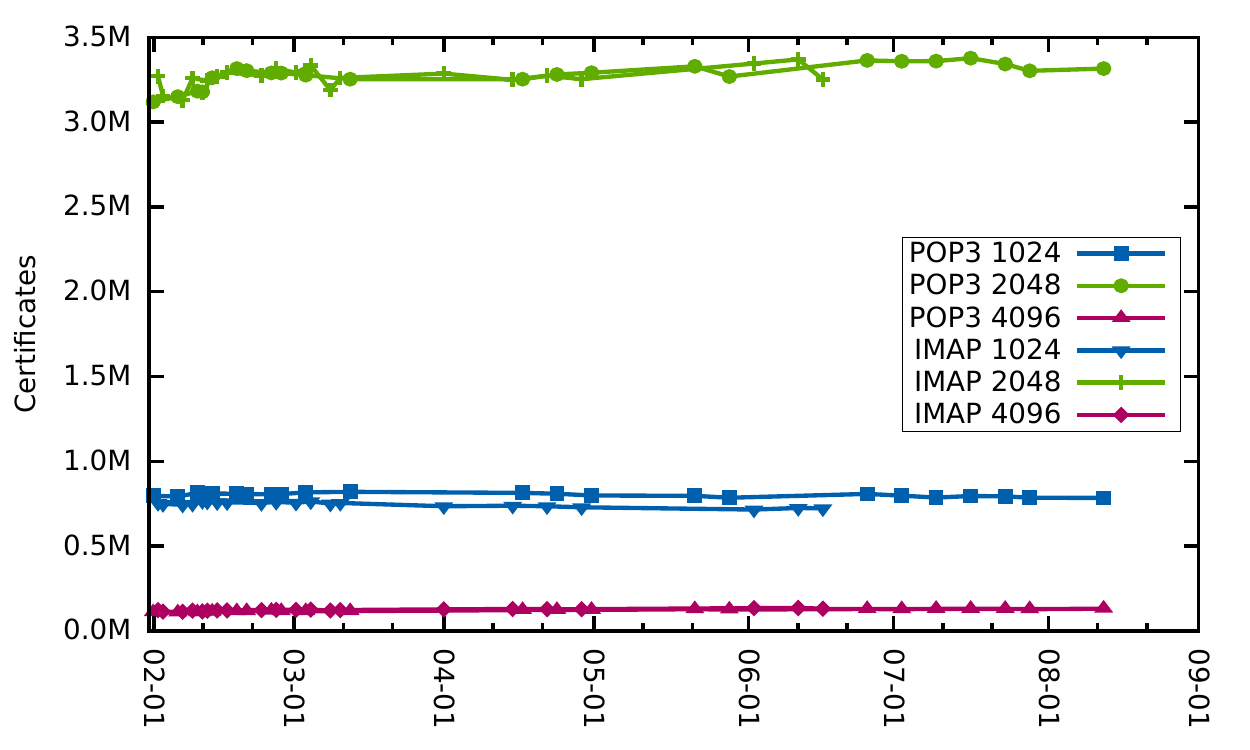}
	\caption{IMAP and POP3 leaf certificate volatility}
	\label{fig:mua_volatility}
\end{figure}

To analyse the change in certificate strength over time we used the freely available datasets from scans.io. They cover the last six months on bi-weekly granularity and are based on scans of the entire IPv4 range. We extracted the size of the RSA key of the leaf certificates and plotted them over time for all important e-mail ports. Figure~\ref{fig:mua_volatility} shows the key distribution of each port for STARTTLS-enabled client-to-server protocols. While the majority uses 2048 bit strength leaf certificates, only a small fraction uses the stronger 4096 bit size. They seem to be rather stable over time, with a slight increase in the overall number of 2048 bit keys. The key distribution for SMTP is shown in Figure~\ref{fig:smtp_volatility}.



\subsection{Domains}
\label{ssec:domains}

Our scans are based on IPv4 addresses. The responsible mail server address for a specific domain is stored in MX records of the domain name system.
The influence of each SMTP server deployment to the global state of TLS security is not evenly distributed, because a single SMTP server can be responsible for many domains and assigned to many different public IP-addresses.
We analyzed MX records for the domains listed in the Alexa Top 1 Million ranking. 
Although this ranking is based on web traffic, we can identify common used SMTP servers.
We checked 978,842 domains and resolved all MX records to IPs.
12.35\% of the domains had no MX record at all.   
From the remaining 857,910 domains 16.01\% used MX records attributed to Google.
\textit{secureserver.net} was included in 4.10\% and \textit{qq.com} in 1.22\% of all domains MX records.
On the contrary, 527,680 MX record names are only used by one domain. 
This gives us a basic idea on how much some players influence the global state of TLS security for SMTP.

Google started to publish statistics of encrypted e-mail traffic as part of their transparency report~\cite{saferEmail}. They state that 45\% of inbound and 20\% of outbound e-mail traffic from Google is still unencrypted. Also, Twitter recently published data stating that only 8\% of their e-mail traffic is unencrypted.

These transparency reports are important sources of information. Due to the decentralized infrastructure of the e-mail transport system, we only get a very limited view of the global state of encryption. More transparency reports would help to clarify our view even more. 
To draw a total picture of SMTP encryption and estimate changes in the usage of different cipher suites we introduce a new approach in Section~\ref{section:cipherIslands}.

\subsection{Plaintext authentication}

A commonly observed issue with e-mail are servers supporting plaintext authentication without support for
STARTTLS. Despite having nothing to do with cryptography per-se, this is an important issue for transport
security as connection authentication data sent via plaintext channel provides an easy target for any attacker.
Further, \texttt{AUTH-PLAIN} should only be offered after a connection has been upgraded via STARTTLS.
If a server offers plaintext authentication on a plaintext connection, neither server nor user will notice
if STARTTLS stripping is performed and password data sent in clear over the wire.
Table~\ref{auth-plain} outlines the number of hosts that offer \texttt{AUTH-PLAIN} and support STARTTLS but
allow plaintext authentication before an upgrade to TLS and hosts that do not offer STARTTLS at all.

\begin{table}[h]
\centering
\begin{tabular}{lrrr}
\hline
Port   & no STARTTLS  & STARTTLS & Total Hosts\\
\hline
25  & 12.90\%         & 24.21\%  & 7,114,171 \\
110 & 4.24\%          & 63.86\%  & 5,310,730 \\
143 & 4.38\%          & 66.97\%  & 4,843,513 \\
587 & 15.41\%         & 42.80\%  & 2,631,662 \\
\hline \\
\end{tabular}
\caption{Hosts that offer \texttt{AUTH PLAIN}}
\label{auth-plain}
\end{table}

\section{Cipher Islands} 
\label{section:cipherIslands}
%

To simulate the consequences of changes in the configured cipher suites between all SMTP servers we have built a model to estimate given probabilities based on the data we collected. This was done to calculate the impact in percent that two random servers are able to communicate using TLS if certain cipher suites are allowed or prohibited. We assume that it is equally likely that a random SMTP relay has the objective to transmit an e-mail for any of the others for the sake of brevity (which is clearly not the case). Since these connections can go unrestrictively in both directions -- i.e. server X can send a message to any server Y and vice versa -- our model still gives useful insights into the consequences of TLS cipher hardening and the estimated use of TLS for specific scenarios as outlined below. To calculate the probabilities we built a weighted graph where the nodes are representing the unique combinations of cipher suites, in total roughly 90,000. The nodes were weighted with the number of SMTP servers that have that specific cipher suite configured. Edges were assigned between two nodes if they share at least one common cipher suite. Table~\ref{tbl:smtp} shows the initial table from the entire set of SMTP servers on port 25 we collected. The list is sorted by percentage and shows that the likelihood for any two servers chosen at random to use a cipher suite in TLS version SSLv3 or TLSv1 is 41\%. In one out of ten cases the servers are unable to agree on a mutual cipher suite, and e-mail will be transmitted in plaintext. Please note that the actual calculations were conducted on each supported cipher suite, and probabilities are only presented in the aggregated TLS version. Probabilities of less than 1\% have been dropped from the table. \\

\begin{table}
    \centering
    \begin{tabular}{lr}
        \hline
        TLS version                &   Probability \\ 
	   \hline
        SSLv3, TLSv1                 &    41.63\%   \\ 
        SSLv3, TLSv1, TLSv1.1, TLSv1.2  &  16.10\%   \\
        TLSv1                       &     14.75\%   \\ 
        Plaintext                   &     10.88\%   \\ 
        TLSv1, TLSv1.1, TLSv1.2       &    8.82\%  \\ 
        SSLv3                       &      3.42\%  \\ 
        TLSv1.2                     &      2.07\%  \\ 
\hline \\
    \end{tabular}
    \caption{Cipher suites in the SMTP dataset}
    \label{tbl:smtp}
\end{table}

Simply changing the cipher suites of a specific server can render large percentages of other SMTP servers unreachable, as they can no longer negotiate a bilateral usable TLS cipher suite. Since the fallback in SMTP is to transmit the e-mail in plaintext if there is no agreement on a shared cipher suite, this is clearly undesirable. Assuming that a considerate administrator changes the cipher suites of her e-mail server according to the recommendations from either \textit{bettercrypto.org}~\cite{bettercrypto} (all A-ranked cipher suites) or RFC7525~\cite{rfc7525}. Since the results for both recommendations are extremely similar, the consequences can be seen in Table~\ref{tbl:smtp:bettercrypto}. It can be seen that more than half of all SMTP servers that support TLS become unreachable as a consequence. \\

\begin{table}
    \centering
    \begin{tabular}{lr}
    \hline
        Cipher suite                         &   Probability \\ 
\hline
        Plaintext                               &     52.80\%  \\ 
        TLS\_ECDHE\_RSA\_AES256\_SHA384     &     18.47\%   \\ 
        TLS\_DHE\_RSA\_AES256\_SHA256       &     11.61\%  \\ 
        TLS\_ECDHE\_RSA\_AES256\_GCM\_SHA384 &    10.59\%  \\ 
        TLS\_DHE\_RSA\_AES256\_GCM\_SHA384   &     6.53\% \\  
\hline \\
    \end{tabular}
    \caption{Cipher suites with SMTP compared to bettercrypto}
    \label{tbl:smtp:bettercrypto}
\end{table}


With the recent discussions whether RC4 is still a secure cipher~\cite{rfc7465} for use in transport security, we also simulated what happens if RC4 is no longer used in any TLS version. The results can be seen in Table~\ref{tbl:smtp:rc4}. As it can be seen the probabilities do not drastically change compared to the initial Table~\ref{tbl:smtp}. About one in ten e-mails would still be transmitted in plain, but the other supported cipher suites substitute the lack of RC4 quite well. \\

\begin{table}
    \centering
    \begin{tabular}{lr}
        \hline
        TLS version                         &   Probability \\ \hline
        SSLv3, TLSv1                 &    40.56\%   \\  
        SSLv3, TLSv1, TLSv1.1, TLSv1.2 &    16.31\%   \\   
        TLSv1                       &    14.41\%    \\   
        Plaintext                       &    11.44\%   \\   
        TLSv1, TLSv1.1, TLSv1.2       &     9.04\%  \\   
        SSLv3                       &     2.99\%  \\   
        TLSv1.2                     &     2.41\%  \\   
        SSLv3, TLSv1.1, TLSv1.2       &     1.10\%  \\   
\hline \\
\end{tabular}
    \caption{Cipher suites without RC4 in SMTP}
    \label{tbl:smtp:rc4}
\end{table}


\section{Discussion} 
\label{section:discussion}
The results from our scans cast a poor light on the use of TLS in e-mail. Not only are weak encryption mechanism and cipher suites like  export grade ciphers supported by a non-negligible fraction of e-mail servers, we also found that the recent increase in HTTPS certificate security (moving certificates from 1024 to 2048 bit) went totally unnoticed for all e-mail related ports, IPv4-wide. Furthermore, millions of hosts are currently misconfigured to allow \texttt{AUTH-PLAIN} over unencrypted connections resulting in the risk of transmitting user credentials unencrypted. Even if each of these systems has only one user, this leaves close to 20 million users vulnerable to attack. \\

Using our model we showed that increasing the cipher suites to the current acceptable level of security would prevent communicating with 50\% of all e-mail servers that support and use TLS, since they are no longer reachable over bilaterally supported TLS. This means that the movement to more secure configurations needs to be done on a larger scale, protecting one server alone is insufficient. Administrators should be encouraged to check with which e-mail servers their systems interact regularly, and adapt their security accordingly. We also showed that the entire e-mail ecosystem as a whole is not dependent on RC4 nor SSLv2 nor SSLv3 alone: disabling them comes with minimal impact, and should be considered for the near-term future. \\

We found that more than 650,000 SMTP deployments accept export grade ciphers with TLSv1.1 and TLSv1.2 (with at least OpenSSL is affected, implementation-wise) although this behaviour is explicitly forbidden (\texttt{MUST NOT}) in RFC 4346~\cite{rfc4346}. \\

We also found troublesome issues despite the cryptographic primitives used (if any at all). More than half of all certificates we observed were self-signed, meaning that an active attacker could exchange the certificate in a man-in-the-middle scenario~\cite{holz2012x}. Since the user interface of most e-mail clients lack a convenient way to check the authenticity of certificates, this leaves clients without the most simple protection mechansims -- which are well-deployed and well understood for the HTTPS PKIX ecosystem. \\

Compared to previous work on HTTPS~\cite{eckersley2010observatory,durumeric2013analysis} we did not analyze the certificate authorities who issued the certificates at length. For one, we found that a very large percentage of certificates in e-mail protocols are self-signed. Secondly, while HTTPS enables the client to authenticate the final server serving the content, this is different for e-mail due to the distributed nature of e-mail transmission protocols. Only the first hop is observable to the client, and thus only a fraction of servers along the path of an e-mail. Lastly, we find this to be rather a policy issue, i.e. who is trusted by the software vendors as certificate authority, then a trust issue, i.e. do I trust CA X to issue certificates for a given domain Y~\cite{soghoian2012interception}. Furthermore, there are now effective countermeasures available for HTTPS which are transparent for users like HPKP~\cite{rfc7469} for dynamic public key pinning or certificate pinning within the browser to effectively prevent man-in-the-middle attacks. No such mechanisms exist so far for e-mail protocols and server implementations, and as such we purposely didn't evaluate the issuing certificate authorities. However, we will release our data sets and the collected certificates for readers who are interested in this analysis.

\subsection{Who leads the way?}
The overall security of the e-mail ecosystem cannot be improved by a single actor. Overall security in e-mail transmission can only improve slowly if the majority of transmitting hosts aligns. We identified multiple issues that need to change not only in the long-term, but already in the near-term future. We found that big providers of e-mail infrastructure often lead the way regarding TLS in e-mail. For one, the usage of TLS in e-mail has increased since the documents by Edward Snowden on NSA dragnet surveillance were published, with Yahoo and Microsoft \textit{live.com} enabling their server infrastructure to enforce TLS wherever possible. Secondly, big service providers like Gmail are in the position to give deep insight into the server-to-server use of TLS. With the release of more and more transparency reports (most recently by Twitter), we can draw a more complete picture of TLS use and identify global actors that handle large volume of e-mail and do not support TLS. Based on this data it would be great to have a list of the most important domains regarding legit e-mail volume, similar to the often-cited Alexa list of popular websites. \\

There will always be space for niche providers like hushmail, riseup.net, or the German Posteo that focus their business model around security and privacy. They target specific privacy-aware users and implement additional security features like DNSSEC, DANE, DKIM, SPF and more. However, with free e-mail giants like live.com, Google, Yahoo and others that have a business model around targeted advertisement depending on the content of their customers e-mails, they still have an implicit responsibility to protect their customers at least by using and enforcing TLS for e-mail wherever possible (and increasingly do so).

\subsection{Limitations}
Since there is no comprehensive list and global ranking of e-mail domains regarding their e-mail volume and overall popularity, this is a clear drawback of our approach. We treated and weighted all scanned IPs equally. Major e-mail provider that handle an overproportional volume of clients are treated equally to the single-user, single-domain self-hosted server. However, we tried to counter this form of bias in our observations and findings by using for one the most popular e-mail related domains as published by Google, as well as correlate it with the most popular websites, even though there seems to be no direct correlation between popular websites and popular e-mail domains. A full list of popular SMTP hosts would have been very helpful, but without any data openly available we had to resort to unweighted statistical analysis. Another limitation is that our findings are only showing a point-in-time snapshot and as such are only of short-living value. 

%

\subsection{Future Work}
For future work we intend to keep our scans actively running and will keep publishing their results as the volatility in cipher configurations and certificates demands regular scan intervals. Long-term trends and reaction times to security incidents like Heartbleed~\cite{durumeric2014heartbleed} or the weak certificates issued with Debian's faulty PRNG~\cite{yilek2009private} back in 2008 are still not fully understood, and data is needed to show the improvements of technical mechanisms to enhance the security in e-mail as well as the awareness of e-mail server administrators. We also plan to keep improving our scanning infrastructure, to allow for long-term data collection and to provide open data for analysis. Additionally, we plan to present our results interactively, on a website which also incorporates existing TLS recommendations. \\

Another aspect we would like to investigate is the issue of fingerprinting: based on the data we collected, how reliable is a software version identifyable just by inspecting the TLS configuration in use. This would include cipher parameters like primes and Diffie-Hellman group in use, as well as default configurations of supported cipher suites.


\section{Mitigation Strategies}
\label{section:mitigation}
Various approaches are technically feasible to increase the security of TLS in e-mail, despite trying to increase the TLS configuration of each and every individual server. The most important issue however is in our opinion to increase the awareness for security and secure configurations among administrators. Furthermore, mitigation can be put into two different categories. For one, various forms of pinning could be used (either on the host itself or on the network layer) as the certificate information is transmitted prior to the encrypted communication and the number of e-mails transmitted to particular servers is usually much larger compared to the number of its certificate changes. \textit{Tack.io}\footnote{\url{https://tack.io}} is a draft from 2013 which could be used to add trust assertions to certificates and would be especially valuable to protocols making use of STARTTLS. \\

The second class of defenses includes all methods where signed certificates are transmitted over additional communication channels, most notably DNSSEC or DANE~\cite{rfc6698}. Both rely on including signed assurances in DNS name records that can be verified by the client to not have been manipulated on the path of communication. Notary services like \textit{ICSI Certificate Notary}~\cite{amann2012extracting}, Convergence\footnote{\url{https://convergence.io}} or the SSL Observatory from the EFF are just a few examples of readily-available and deployed approaches where the trust is distributed among independent entities. Defending against active attackers is much harder however, as the attacker is always in the position to suppress communication like the STARTTLS command at the beginning of the transmission. Such stripping attacks can only be prevented if there is no fallback mechanism to plaintext, which is currently unfortunately the default between MTAs. In IETF a new draft is being worked on called DEEP (``Deployable Enhanced Email Privacy'')\footnote{https://datatracker.ietf.org/doc/draft-ietf-uta-email-deep}. If standardized and implemented it would offer extensions to SMTP, POP3 and IMAP for so-called ``latches'' which are similar to what HSTS~\cite{rfc6797} does for HTTP, follow-up connections will use the same or better security parameters as previous transactions with a given server. The extensions also offer mail user agents (MUA) the possibility to indicate security levels to end-users. \\


Lastly, operators of smaller e-mail servers often do not change default configurations. They should be incentivized to enable important security primitives, or enforcing TLS for communication. Guidelines to improve security for various software products exist~\cite{bettercrypto,rfc7525,MozillaTLSrecommendations}, but these suggestions are not automatically implemented. If software products embed secure configurations by default, this could change the overall primitives in TLS usage. Smart update mechanisms could enforce, or at least offer, the use of stronger cryptographic primitives. A promising project for exterminating the usage of self-signed certificates is the project \textit{Let's encrypt}\footnote{\url{https://letsencrypt.org/}} which will become a fully automatic CA to issue trusted certificates for free.

\section{Related Work}
\label{section:relatedWork}
\subsubsection{Large-scale Scanning}
In 2008 Heidemann et~al.~\cite{heidemann2008census} started an Internet wide discovery scan of edge hosts. They were the first to census the IPv4 address space, but were limited to host discovery, not containing protocol or service specific results. 
First projects to cover TLS related scanning included the EFF (Electronic Frontier Foundation) Observatory~\cite{eckersley2010observatory,eckersley2010ssliverse} which collected 1.3 million unique certificates of the HTTPS / TLS public key infrastructure. Holz et~al.~\cite{holz2011ssl} verified flaws in the PKI of TLS with a larger data set obtained by using nmap on the Alexa Top 1 Million pages as well as passive scan data. Amann~et~al.~\cite{amann2012revisiting} revisited SSL and X.509 deployments in 2012 with passively collected data. In 2013 Durumeric et al. developed zmap, a tool for fast Internet-wide scanning~\cite{durumeric2013zmap}.  
This resulted in the HTTPS ecosystem study~\cite{durumeric2013analysis}, on certificates and certificate authorities, having a nearly complete dataset. 
These studies are the inspiration for our work, but primarily focused on the public key infrastructure of HTTPS. Unlike our study, detailed cipher suite scans were not conducted. The results of port scans originated from University of Michigan were used as input data for our scan.

\subsubsection{Evaluating HTTPS}
In 2007 Lee et~al.~\cite{lee2007cryptographic} surveyed 19,000 TLS enabled servers and inspected the supported TLS versions and different cryptographic primitves for HTTPS servers. Several newer studies for different deployed security mechanisms exist. Huang et~al.~\cite{huang2014experimental} surveyed forward secrecy parameters and measured latencies of different cipher suites, Kranch and Bonneau~\cite{kranch2015upgrading} conducted a study of HTTP Strict Transport Security and public key pinning deployments. Weissbacher~et~al.~\cite{weissbacher2014csp} analyzed the adoption of CSP (Content Security Policy). These datasets are all limited to the Alexa Top 1 Million ranking and, especially in case of CSP and HSTS, not relevant for e-mail security. The acceptance of cipher suites which support forward secrecy is analyzed in our study.
Heninger~et~al.~\cite{heninger2012mining} performed Internet wide scans to identify vulnerable RSA and DSA keys. Due to entropy problems of the random number generators used, they were able to obtain 0.5\% of all RSA private keys for HTTPS. We revisited their work and analyzed RSA public keys used in TLS over e-mail related protocols. Zhang~et~al.~\cite{zhang2013mismanagement} used one scan of the HTTPS ecosystem scans to find untrusted certificates. These certificates were used as a metric, among others, to find a relationship between misconfiguration and malicious activity of networks. Adrian~et~al.~\cite{adrian2015imperfect} investigated the security of Diffie-Hellman key exchanges using an Alexa Top 1 Million scan. This scan was also performed for HTTPS, not considering the e-mail ecosystem. We investigated common primes and Diffie-Hellman parameters, as well as usage statistics on elliptic curve Diffie-Hellman exchanges for the e-mail ecosystem.

\subsubsection{Scan Detection}
Dainotti~et~al.~\cite{dainotti2012analysis} measured the impact of a coordinated Internet wide scan using the UCSD network telescope, a /8 darknet receiving unsolicited traffic. 
Also Durumeric~et~al.~\cite{durumeric2014internet} observed Internet-wide scanning activity, identified patterns and analyzed defensive behaviour.

\section{Conclusion}
\label{section:conclusion}
In this paper we showed and evaluated a scalable approach to assess the overall security in e-mail protocols by inspecting the underlying TLS primitives. We evaluated our methodology on all different ports in use for e-mail by actively scanning the entire range of IPv4. We conducted more then 10 billion TLS handshakes and discovered multiple flaws that prevent the use of TLS as effective countermeasure against passive attackers. Overall, the state of TLS in e-mail transmission is, unsurprisingly, worse than compared to HTTPS, as 15--30\% of all servers accept weak export grade ciphers, and the majority of certificates is self-signed. On the positive side we were able to show that RC4 and the use of SSLv2 and SSLv3 for backward compatibility can be considered almost obsolete. Disabling them has hardly any negative impact on the total number of reachable SMTP servers according to our calculations.

\section*{Acknowledgements}
We would like to thank the HPC team at the Gregor Mendel Institute which allowed us to use MENDEL for parts of the evaluation. Thanks to Hanno Böck for discussion and feedback on Cryptography related subjects, RSA Weak-Keys and TLS. Thanks to Viktor Dukhovni for valuable feedback and insights into OpenSSL specifics, MTA operational details and discussion on opportunistic e-mail encrytion in general. Thanks to Wolfgang Breyha for discussion on applied e-mail security.

The research was funded by COMET K1, FFG - Austrian Research Promotion Agency and under grant 846028.​

\bibliographystyle{abbrv}
\bibliography{bibliography_httpsScanning,rfc}

\end{document}